\documentclass[fleqn,twoside]{article}
\usepackage{espcrc2}

\usepackage{epsfig}
\usepackage[figuresright]{rotating}

\def\bd{B^0}

\def\bs{B^0_s}

\def\bmix{B^0 \mbox{--} \overline{B^0}}

\usepackage{relsize}
\def\babar{\mbox{\slshape B\kern-0.1em{\smaller A}\kern-0.1em
    B\kern-0.1em{\smaller A\kern-0.2em R}}}
\def\B     {\ensuremath{B}}
\def\Bbar  {\kern 0.18em\overline{\kern -0.18em B}{}}
\def\BB    {\ensuremath{B\Bbar}} 
\def\Bz    {\ensuremath{B^0}}

\def\BF    {\ensuremath{{\cal B}}}

\def\Bs    {\ensuremath{B^0_s}}

\def\ellell     {\ensuremath{\ell^+ \ell^-}}
\def\epem       {\ensuremath{e^+e^-}}
\def\etap       {\ensuremath{\eta^{\prime}}}
\def\invfb   {\ensuremath{\mbox{\,fb}^{-1}}}
\def\KS    {\ensuremath{K^0_{\scriptscriptstyle S}}} 
\def\KL    {\ensuremath{K^0_{\scriptscriptstyle L}}} 
\def\Kbar  {\kern 0.2em\overline{\kern -0.2em K}{}}
\def\Kstarz  {\ensuremath{K^{*0}}}

\def\Kstar   {\ensuremath{K^*}}

\def\Kstarp   {\ensuremath{K^{*+}}}

\def\Kzb   {\ensuremath{\Kbar^0}}
\def\gev  {\ensuremath{\rm \,Ge\kern -0.08em V}} 
\def\gevc {\ensuremath{{\rm \,Ge\kern -0.08em V\!/}c}} 
\def\mev  {\ensuremath{\rm \,Me\kern -0.08em V}} 
\def\mevc {\ensuremath{{\rm \,Me\kern -0.08em V\!/}c}} 
\def\piz   {\ensuremath{\pi^0}}
\def\pizs  {\ensuremath{\pi^0\mbox\,\rm{s}}}
\def\qqbar {\ensuremath{q\overline q}}

\mathchardef\Upsilon="7107
\def\Y#1S{\ensuremath{\Upsilon{(#1S)}}\xspace}
\def\FourS {\ensuremath{\Upsilon{(4S)}}}

\newcommand{\AmS}{{\protect\the\textfont2
  A\kern-.1667em\lower.5ex\hbox{M}\kern-.125emS}}

\title{Rare Hadronic and Radiative Penguin B Decays at \babar}
\author{S. Willocq\address[UMASS]{Physics Department,
        University of Massachusetts, Amherst, MA 01003, USA}
        ~representing the \babar\ Collaboration}

\begin{document}

\pagestyle{empty}


\begin{flushright}
{\normalsize
\babar--CONF--01/99\\
SLAC--PUB--9139\\
February, 2002\\}
\end{flushright}

\vspace{20mm}

\begin{center}

{\large\bf Rare Hadronic and Radiative Penguin B Decays at \babar}

\vspace{20mm}

St\'ephane Willocq \\
Department of Physics \\
University of Massachusetts, Amherst, MA 01003, USA \\ ~~\\
Representing the \babar\ Collaboration \\

\end{center}

~~~
\vspace{10mm}
~~~
\hfill\break

\begin{center}
{\bf Abstract}
\end{center}
\vspace{2mm}

{\normalsize
  We report recent results in the study of rare hadronic and
radiative penguin decays of $B$ mesons. These results are based
on a sample of 23 million \BB\ pairs collected by the \babar\
Collaboration at the SLAC PEP-II \epem\ B Factory.
}
\begin{center}

\vspace{25mm}
{\normalsize
     Contributed to the Proceedings of the 5$^{th}$ KEK Topical Conference:\\
     Frontiers In Flavor Physics (KEKTC5),\\
     20-22 Nov 2001, Tsukuba, Ibaraki, Japan.}

\vspace{20mm}
{\sl Stanford Linear Accelerator Center, Stanford University, Stanford, CA 94309}\\
\vspace{1mm}
\hrule
\vspace{1mm}
Work supported in part by Department of Energy contract
                DE-AC03-76SF00515.\\
\end{center}

\pagebreak
\pagebreak

\begin{abstract}

  We report recent results in the study of rare hadronic and
radiative penguin decays of $B$ mesons. These results are based
on a sample of 23 million \BB\ pairs collected by the \babar\
Collaboration at the SLAC PEP-II \epem\ B Factory.

\vspace{1pc}

\end{abstract}


\maketitle

\section{MOTIVATION}

  A major effort is currently underway to test the flavor sector of
the Standard Model (SM) in general and the origin of CP violation
in particular.
Measurements of the time-dependent CP-asymmetries in
$B^0 \to J/\psi K^0_s$ decays have led to the first observation of
CP violation~\cite{sin2b} in the $\bmix$ system and to a measurement
of the phase $\beta$ of the Cabibbo-Kobayashi-Maskawa (CKM)
matrix element $V_{td}$--the current world average is
$\sin 2\beta = 0.79 \pm 0.10$.
The magnitude of $V_{td}$ is constrained by studies of $\bd$ and $\bs$ mixing
and is expected to be determined with an uncertainty of $\sim 5\%$
in the near future.

  The determination of the magnitude and phase ($\gamma$) of the CKM element
$V_{ub}$ is expected to be much more difficult, due in part to the small
rate for $b \to u$ transitions ($\sim 0.5 - 1\%$
compared to the rate for $b \to c$ transitions).
In parallel with the measurement of $\sin 2\beta$ with $B^0 \to J/\psi K^0_s$
decays, one can in principle determine $\sin 2\alpha$ from the
time-dependent asymmetry in $B^0 \to \pi^+ \pi^-$ decays
(where $\alpha = \pi - (\beta + \gamma)$).
However, the study of charmless two-body $B$ decays has shown that
gluonic penguin $b \to d$ processes contribute significantly to the
overall $B^0 \to \pi^+ \pi^-$ decay rate and have
a different weak phase than the dominant $b \to u$ tree process,
thereby preventing a straightforward interpretation of the
measured asymmetry.
Improved understanding of gluonic penguins is thus of very high importance
and constitutes an important challenge for current experiments.

\break
  Another motivation to study rare hadronic and radiative penguin decays
is that many of these decays proceed via effective flavor-changing
neutral current (FCNC) transitions. Such FCNC processes are forbidden
at tree level and involve heavy particles in loops (e.g. $W$ boson
or top quark). This implies that the rates are low but also that such
processes are sensitive to the presence of heavy non-SM particles in the loops.

  Besides the measurement of decay rates, one can also study CP violation in
the decay (a.k.a. direct CP violation). In this case, the relevant
observable is
\begin{equation}
  A_{CP}   =   \frac{\Gamma(\overline{B} \to \overline{f})-\Gamma(B \to f)}
                    {\Gamma(\overline{B} \to \overline{f})+\Gamma(B \to f)}\: .
  \label{equ:acpdef}
\end{equation}
If both penguin and tree decay amplitudes contribute to the
decay, the asymmetry can be expressed as
\begin{equation}
  A_{CP}   =   \frac{2 |P| |T| \sin\Delta\phi \: \sin\Delta\delta}
               {|P|^2 + |T|^2 + 2 |P| |T| \cos\Delta\phi \: \cos\Delta\delta}
  \: ,
  \label{equ:acptp}
\end{equation}
where $|P|$ and $|T|$ represent the magnitudes of the penguin and tree decay
amplitudes, whereas $\Delta\phi$ and $\Delta\delta$ correspond to the
difference between the weak and strong phases for the penguin and tree
amplitudes, respectively.
From this expression it can be seen that $A_{CP}$ is expected to be very small
in the SM (less than 1\%) for penguin-dominated decays like
$B \to \phi K^{(\ast)}, K^0 \pi, K^\ast \gamma$, and thus the observation
of large CP asymmetries in these modes
would clearly signal the presence of new physics
in the penguin loops.

\section{EXPERIMENTAL DETAILS}

In the following, we report the results of analyses
based on a sample of 23 million
\BB\ events collected by the \babar\ Collaboration at the
asymmetric-energy PEP-II \epem\ collider.
This sample corresponds to 20.7 \invfb\ of \epem\ annihilation data
recorded at the \FourS\ resonance.
An additional (off-resonance) sample corresponding
to 2.6 \invfb\ recorded 40 \mev\
below the resonance is also used to study continuum backgrounds.
Charged particles are tracked and measured with the
combination of a five-layer double-sided
silicon microstrip detector and a 40-layer drift chamber
immersed in a 1.5 T solenoidal magnetic field.
Surrounding the tracking systems, a Cherenkov ring imaging
detector (DIRC) relying on
total internal reflection of the Cherenkov light produced
in 4.9 m-long quartz bars provides charged particle identification.
A Thallium-doped CsI electromagnetic calorimeter is used
for photon detection and electron identification.
Finally, an instrumented flux return provides muon and
\KL\ identification.
Further detail about the \babar\ detector is given
elsewhere~\cite{bbrdet}.

Most of the analyses presented here attempt to fully reconstruct the
\B\ decay. To do so, we rely on the well-known beam energies and
exploit kinematical constraints from energy and momentum
conservation in the $\epem \to \FourS \to \BB$ process.
Two mostly uncorrelated kinematical variables are used:
the beam-energy substituted mass,
$m_{ES} = \sqrt{E_{beam}^{\ast 2} - {\bf p}_B^{\ast 2}}$,
and the energy difference in the \epem\ center-of-mass system (CMS),
$\Delta E = E_B^\ast - E_{beam}^\ast$,
where $E_{beam}^\ast$ and $E_B^\ast$ are the energies of the
beam and the \B\ meson candidate, and
${\bf p}_B^\ast$ is the \B\ meson candidate momentum vector,
all quantities being defined in the CMS.
In some analyses, a kinematical fit is performed
on the momenta of the \B\ decay products by imposing
$E_B^\ast = E_{beam}^\ast$ and the corresponding
energy-constrained mass $m_{EC}$ is used.
The $m_{ES}$ resolution is dominated by the beam energy
spread and has a typical value of 2.5 \mev.
The $\Delta E$ resolution is different for each particular
final state and ranges between 20 and 70 \mev\
(the resolution is noticeably worse for modes containing
photons or \pizs).

With the exception of the $B \to K^{(\ast)} \ellell$ analysis,
the background from \BB\ events is small or negligible for
all other analyses.
The dominant background from $\epem \to \qqbar$ continuum
events ($q = u, d, s, c$) is suppressed using event shape
and energy flow variables. These variables exploit the fact
that \BB\ events are much more spherical than continuum events
due to the small \B\ momentum ($\sim 340$ \mevc) in the CMS.

Analyses have been optimized and tested using
signal and background Monte Carlo samples, as well as
off-resonance data and side-band data. The signal region
remained hidden during this optimization and validation process.

\section{RADIATIVE PENGUIN DECAYS}

  As mentioned earlier, the measurement of \Bz\ and \Bs\ oscillation
frequencies will soon provide a determination of the
ratio $|V_{td} / V_{ts}|$ with an uncertainty of $\sim 5\%$.
Radiative penguin decays are also sensitive to the same CKM elements
and can provide a complementary determination via the ratio
$\Gamma(B \to \rho \gamma) / \Gamma(B \to \Kstar \gamma)$.

  A first step toward such a measurement is the determination of
the branching fraction for $B \to \Kstar \gamma$ decays.
The analysis proceeds with the selection of high-energy photons
with $1.5 < E_\gamma < 4.5 \gev$ and
$2.30 < E_\gamma^\ast < 2.85 \gev$ in the laboratory and CMS
frames, respectively.
Photons consistent with originating from \piz\ and $\eta$ decays
are removed.
Photons are further required to be isolated to remove
contamination from high-energy \piz\ decays.
\Kstar\ candidates are reconstructed in $K^+ \pi^-$, $\KS \piz$,
$K^+ \piz$ and $\KS \pi^+$ modes\footnote{Charged conjugate states
are implied throughout, unless noted otherwise.},
making use of DIRC information
to select charged kaons.
The dominant backgrounds consist of events with photons from
initial state radiation or from \piz\ and $\eta$ decay. These
are suppressed by utilizing event shape variables, the polar angle
of the \B\ candidate and the helicity angle of the \Kstar\ decay.
 
\begin{figure}[t]
  \hspace{-0.4cm}
  \epsfxsize8cm
  \epsfbox{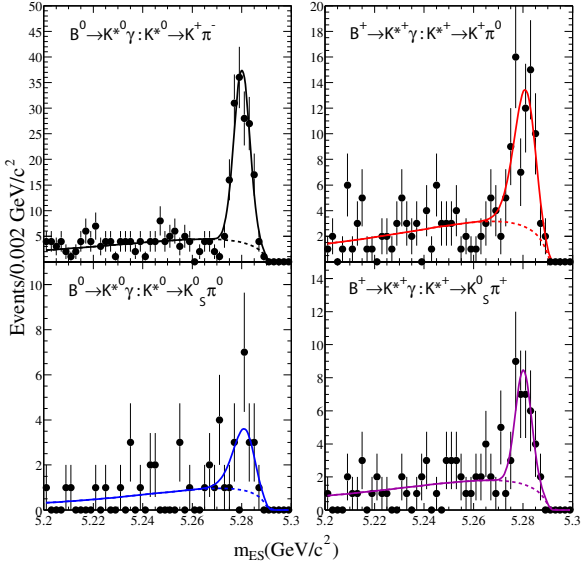}
\vspace{-1.2cm}
\caption{Distribution of $m_{ES}$ for $B \to \Kstar \gamma$ candidates
  in four different \Kstar\ decay channels.}
\label{fig:kstargamma}
\end{figure}

  $B \to \Kstar \gamma$ candidates are required to satisfy
$m_{ES} > 5.2 \gev$ and $-0.2 < \Delta E < 0.1 \gev$
(for $K^+ \pi^-$ and $\KS \pi^+$ modes)
or $-0.225 < \Delta E < 0.125 \gev$
(for $\KS \piz$ and $K^+\piz$ modes).
An unbinned likelihood method is used to extract signal yields from the
$m_{ES}$ distributions, see Fig.~\ref{fig:kstargamma}.
The branching fractions for each of the four modes are shown in
Table~\ref{tab:bfracradpeng}.
Averaging the two \Kstarz\ and \Kstarp\ modes yields
$\BF (B^0 \to \Kstarz \gamma)
  = (42.3 \pm 4.0({\rm stat}) \pm 2.2({\rm syst})) \times 10^{-6}$
and 
$\BF (B^+ \to \Kstarp \gamma)
  = (38.3 \pm 6.2({\rm stat}) \pm 2.2({\rm syst})) \times 10^{-6}$.
These values, although somewhat lower, agree with recent
theoretical predictions, given the $\sim 40\%$ uncertainty in
the predictions.

  We have also searched for the rare decay $B^0 \to \gamma\gamma$,
which is predicted to have a branching fraction in the range of
$(0.1 - 2.3) \times 10^{-8}$ within the SM.
However, this could be enhanced by new physics contributing to the
loops. 
Given the current size of the event sample, any observation
would be clear evidence for physics beyond the SM.
The analysis proceeds in a way similar to the $\Kstar \gamma$
analysis. One event is found in the signal region,
defined by the $\pm 2\sigma$
ranges $|m_{ES} -m_{B^0}| < 0.0078 \gev$
and $ |\Delta E| < 0.144 \gev$,
in agreement with the expected background of 0.9 event.
As a result, a limit is set to be
$\BF (B^0 \to \gamma\gamma) < 1.7 \times 10^{-6}$ at the 90\% C.L.
This represents a factor of 20 improvement over the best previous
limit.

  We have searched for $B^+ \to K^{(\ast)+} \ellell$
with $\Kstarp \to \KS \pi^+$ and
$B^0 \to K^{(\ast)0} \ellell$ with $\Kstarz \to K^+ \pi^-$.
Not only does this analysis need to suppress background
from continuum events, it also
needs to suppress background from \BB\ events.
Major sources of \BB\ background are events in which both
\B\ mesons decay semileptonically or one decays to
$J/\psi K^{(\ast)}$ or $\psi({\rm 2S}) K^{(\ast)}$ final states
with $J/\psi \to \ellell$ or $\psi({\rm 2S}) \to \ellell$.
It is particularly important to suppress \B\ decays to
charmonium since these have characteristics that are
similar to the signal.
The event yields are extracted with an unbinned likelihood
fit in the $\Delta E$-$m_{ES}$ plane.
The overall efficiencies for the different modes
range between 6\% for $\Kstarp \mu^+ \mu^-$
and 18\% for $K^+ e^+ e^-$.
No significant signal is observed but the limits
(see Table~\ref{tab:bfracradpeng})
are close to SM calculations.

\begin{table}[tbp]
\caption{Summary of signal yields and branching fraction measurements
         for radiative penguin decays.
         The first and second uncertainties in the branching fractions
         are statistical and systematic, respectively.
         The uncertainty in the yield is statistical only.}
\label{tab:bfracradpeng}
\begin{center}
\begin{tabular}{lcc}
Mode                         & Signal yield         & \BF\ $(10^{-6})$ \\
\hline
$\Kstarz_{K^+\pi^-} \gamma$~\cite{KstrGam}
                             & $135.7 \pm 13.3$     & $42.4 \pm  4.1 \pm 2.2$ \\ 
$\Kstarz_{\KS\pi^0} \gamma$  & $ 14.8 \pm  5.6$     & $41.0 \pm 17.1 \pm 4.2$ \\ 
$\Kstarp_{\KS\pi^+} \gamma$  & $ 28.1 \pm  6.6$     & $30.1 \pm  7.6 \pm 2.1$ \\ 
$\Kstarp_{K^+\pi^0} \gamma$  & $ 57.6 \pm 10.4$     & $55.2 \pm 10.7 \pm 3.8$ \\ 
\hline
$\gamma\gamma$~\cite{GamGam} & ---                  & $< 1.7$ (90\% C.L.) \\
\hline
$K\ellell$~\cite{Kll}        & $  0.0 \pm  0.3$     & $< 0.6$ (90\% C.L.) \\
$\Kstar\ellell$              & $  0.7 \pm  1.1$     & $< 2.5$ (90\% C.L.) \\
\hline

\end{tabular}
\end{center}
\end{table}

\section{RARE HADRONIC B DECAYS}

  Rare hadronic decays have been searched for in two-body,
quasi two-body and three-body modes.
In the case of two-body decays, the following
topologies are examined:
$h^+ h^{\prime -}$, $h^+ \pi^0$, $\KS h^+$, $\KS \pi^0$ and $\KS \KS$,
where $h$ represents a charged pion or kaon.
Signal yields are extracted using an extended likelihood function
taking all relevant topologies into account.
The likelihood incorporates information from event kinematics
($\Delta E$ and $m_{ES}$), event shape and energy flow via a Fisher
discriminant, as well as mode-specific information like the
Cherenkov angle for charged hadrons,
the invariant mass for $\KS \to \pi^+ \pi^-$ candidates or
the helicity angle.
This analysis benefits from the excellent $\pi/K$ separation
provided by the DIRC.
The significance of the separation is greater than
$4\:\sigma$ up to laboratory momenta of 3 \gevc\ and is
$2.5\:\sigma$ at the highest momentum ($\sim 4.3$ \gevc).

  Signal yields and branching fractions are summarized
in Table~\ref{tab:bfrachad}.
The large rate for $K \pi$ final
states as compared to $\pi \pi$ final states confirms
the previous observation by the CLEO Collaboration.
Since tree diagrams for $B^0 \to K^+ \pi^-$ decays are
Cabibbo-suppressed as compared to $B^0 \to \pi^+ \pi^-$
decays, the larger rate for $K \pi$ final states is
a clear indication that gluonic penguin diagrams contribute
significantly to the decay.

\begin{figure}[htb]
  \hspace{-0.4cm}
  \epsfxsize8cm
  \epsfbox{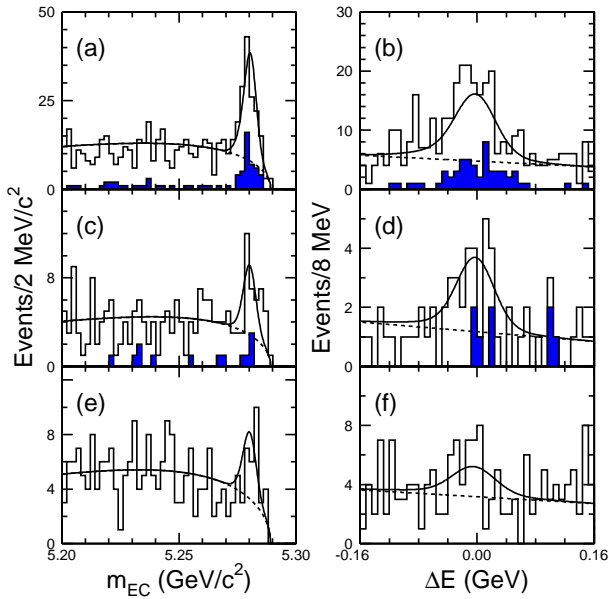}
\vspace{-0.8cm}
\caption{Distributions of $m_{EC}$ and $\Delta E$ for
 (a,b) $B^+ \to \etap K^+$, (c,d) $B^0 \to \etap K^0$,
 and (e,f) $B^+ \to \omega \pi^+$.
 The shaded area corresponds to $\etap \to \eta\pi\pi$.
 The solid curves represent the result of the likelihood function
 and the dashed curves correspond to the background contribution.}
\label{fig:etapom}
\end{figure}

  A host of quasi two-body decays were also studied.
These involve $\omega$, $\eta$, $\etap$ or $\phi$ resonances
and offer another window to study decays mediated
by either or both tree and penguin decay amplitudes.
For example, $B \to \omega \pi$ decays are
dominated by tree decay amplitudes, whereas
$B \to \phi K$ decays are dominated by gluonic
penguin decay amplitudes.
The later decay mode also holds the promise of
a $\sin 2\beta$ measurement via the time-dependent
asymmetry in $B^0 \to \phi \KS$.
Such a measurement is of interest to test the consistency
of the CKM picture in two different decay mechanisms,
$B^0 \to J/\psi K^0_s$ involving $b \to c\bar{c}s$ transitions
and $B^0 \to \phi \KS$ involving $b \to s\bar{s}s$ transitions.
$B \to \phi K^0$ and $B^+ \to \phi \Kstarp$ decays were first observed
by the \babar\ Collaboration.
A significant signal is also observed
in $B^+ \to \omega \pi^+$ decays,
see Table~\ref{tab:bfrachad} and Fig.~\ref{fig:etapom}.

  Previous studies of $B \to \eta^{(\prime)} K^{(\ast)}$
decays reported a surprizingly large branching fraction.
With a larger data sample, we confirm higher than expected
rates for $B \to \eta \Kstar$ and $B \to \etap K$ decays,
see Table~\ref{tab:bfrachad} and Fig.~\ref{fig:etapom}.
Since tree contributions are Cabibbo-suppressed one
needs to invoke a large enhancement in the penguin contributions,
possibly originating from interference between different
penguin amplitudes ($g^\ast \to u\bar{u}$ and $g^\ast \to s\bar{s}$).
 
\begin{table}[tbp]
\caption{Summary of signal yields and branching fraction measurements
         for rare hadronic decays.
         The first and second uncertainties in the branching fractions
         are statistical and systematic, respectively.
         The uncertainty in the yield is statistical only.}
\label{tab:bfrachad}
\begin{center}
\begin{tabular}{lcc}
Mode                         & Signal yield         & \BF\ $(10^{-6})$ \\
\hline
$\pi^+\pi^-$~\cite{2body}    & $ 41   \pm 10  $     & $ 4.1 \pm  1.0 \pm 0.7$ \\
$K^+\pi^-$                   & $169   \pm 17  $     & $16.7 \pm  1.6 \pm 1.3$ \\
$K^+ K^-$                    & $8.2^{+7.8}_{-6.4}$  & $<2.5$ (90\% C.L.) \\
$\pi^+\pi^0$                 & $ 37   \pm 14  $     & $<9.6$ (90\% C.L.) \\
$K^+\pi^0$                   & $ 75   \pm 14  $     & $10.8^{+2.1}_{-1.9} \pm 1.0$ \\
$K^0\pi^+$                   & $ 59^{+11}_{-10}$    & $18.2^{+3.3}_{-3.0} \pm 2.0$ \\
$\Kzb K^+$                   & $-4.1^{+4.5}_{-3.8}$ & $<2.4$ (90\% C.L.) \\
$K^0\pi^0$                   & $17.9^{+6.8}_{-5.8}$ & $8.2^{+3.1}_{-2.7} \pm 1.2$ \\
$K^0 \Kzb$~\cite{kzkzb}      & $3.4^{+3.4}_{-2.4}$  & $<7.3$ (90\% C.L.) \\
\hline
$\phi K^+$~\cite{phiK}       & $31.4^{+6.7}_{-5.9}$ & $7.7^{+1.6}_{-1.4} \pm 0.8$ \\
$\phi K^0$                   & $10.8^{+4.1}_{-3.3}$ & $8.1^{+3.1}_{-2.5} \pm 0.8$ \\
$\phi \Kstarp$               & ---                  & $9.7^{+4.2}_{-3.4} \pm 1.7$ \\
~~~$\phi \Kstarp_{K^+\pi^0}$ & $7.1^{+4.3}_{-3.4}$  & $12.8^{+7.7}_{-6.1} \pm 3.2$ \\
~~~$\phi \Kstarp_{K^0\pi^+}$ & $4.4^{+2.7}_{-2.0}$  & $8.0^{+5.0}_{-3.7} \pm 1.3$ \\
$\phi \Kstarz$               & $20.8^{+5.9}_{-5.1}$ & $8.7^{+2.5}_{-2.1} \pm 1.1$ \\
$\phi \pi^+$                 & $0.9^{+2.1}_{-0.9}$  & $<1.4$ (90\% C.L.) \\
\hline
$\omega K^+$~\cite{etapom}   & $6.4^{+5.6}_{-4.4}$  & $<4$ (90\% C.L.) \\
$\omega K^0$                 & $8.1^{+4.6}_{-3.6}$  & $<13$ (90\% C.L.) \\
$\omega\pi^+$                & $27.6^{+8.8}_{-7.7}$ & $6.6^{+2.1}_{-1.8} \pm 0.7$ \\
$\omega\pi^0$                & $-0.9^{+5.0}_{-3.2}$ & $<3$ (90\% C.L.) \\
\hline
$\eta\Kstarz$~\cite{etak}    & $ 20.5 \pm  6.0$     & $19.8^{+6.5}_{-5.6} \pm 1.7$ \\
$\eta\Kstarp$                & $ 14.3 \pm  6.6$     & $<33.9$ (90\% C.L.) \\
\hline
$\etap K^+$~\cite{etapom}    & ---                  & $70 \pm 8 \pm 5$ \\
~~~$\etap_{\eta\pi\pi} K^+$  & $49.5^{+8.1}_{-7.3}$ & $63^{+10}_{-9}$ \\
~~~$\etap_{\eta\gamma} K^+$  & $87.6^{+13.4}_{-12.5}$ & $80^{+12}_{-11}$ \\
$\etap K^0$                  & ---                  & $42^{+13}_{-11} \pm 4$ \\
~~~$\etap_{\eta\pi\pi} K^0$  & $6.3^{+3.3}_{-2.5}$  & $28^{+15}_{-11}$ \\
~~~$\etap_{\eta\gamma} K^0$  & $20.8^{+7.4}_{-6.5}$ & $61^{+22}_{-19}$ \\
$\etap \pi^+$                & ---                  & $<12$ (90\% C.L.) \\
~~~$\etap_{\eta\pi\pi} \pi^+$& $5.7^{+3.8}_{-2.8}$  & $7.1^{+4.8}_{-3.5}$ \\
~~~$\etap_{\eta\gamma} \pi^+$& $-0.9^{+7.8}_{-6.2}$ & $-0.7^{+6.7}_{-5.3}$ \\
\hline
$a_0^\pm \pi^\mp$~\cite{a0pi}& $18.1^{+8.7}_{-7.4}$ & $<11.5$ (90\% C.L.) \\
\hline
$\rho^\pm\pi^\mp$~\cite{3pi} & $ 89   \pm 16  $     & $28.9 \pm  5.4 \pm 4.3$ \\
$\rho^0 \pi^0$               & $  6.1 \pm  5.8$     & $<10.6$ (90\% C.L.) \\
\hline
$\Kstarz\pi^+$~\cite{Kstrpi} & $ 34.8 \pm  7.6$     & $15.5 \pm 3.4 \pm 1.5$ \\
\hline

\end{tabular}
\end{center}
\end{table}

\break
  To further explore the large decay rate into \etap, we have
performed a semi-inclusive reconstruction of the decay
$B \to \etap X_s$, where $X_s$ represents a hadronic system with
non-zero strangeness.
The \etap\ candidates are reconstructed in the $\eta\pi^+\pi^-$ channel
(with $\eta \to \gamma\gamma$) and are required to have momentum
$p^\ast > 2$ \gevc\ in the CMS to remove most of the
$b \to c \to \etap$ background.
The hadronic system is reconstructed in 16 different modes including
one kaon and at most 3 pions (and no more than 1 $\pi^0$).
To discriminate between all combinations in a given event,
the combination with the lowest value of
$\chi^2 = (\frac{m_{ES}-m_B}{\sigma_{m_{ES}}})^2
        + (\frac{\Delta E}{\sigma_{\Delta E}})^2$
is selected.
The rate for $B \to \etap X_s$ events with
$p_{\etap}^\ast > 2$ \gevc\ is found to be
$(6.8^{+0.7}_{-1.0} \pm 0.6 ^{+0.0}_{-0.5}) \times 10^{-4}$,
where the first error is statistical, the second is systematic,
and the third corresponds to the uncertainty in the contribution
from $B \to \etap D^{(\ast)0}$ decays.
This measurement confirms the large rate previously measured by CLEO.
Such a large rate is generally believed to be due to the
anomalous coupling of the \etap\ to two gluons.

  We have also investigated \B\ decays to three-body final states.
Particularly promising is the decay $B^0 \to \pi^+ \pi^- \pi^0$,
which could provide a clean measurement of the CP-violating
phase $\alpha$ via a Dalitz amplitude analysis.
Due to the small expected signal yield, the current version of the analysis
aims to determine the \B\ decay rates to
$\rho^\pm(770) \pi^\mp$, $\rho^0(770) \pi^0$,
$\rho^\pm(1450) \pi^\mp$, $\rho^0(1450) \pi^0$,
$({\rm charged~scalar})^\pm \pi^\mp$, $f^0 \pi^0$,
and non-resonant $\pi^+ \pi^- \pi^0$ final states.
A significant signal is found only in the $B^0 \to \rho^\pm(770) \pi^\mp$ mode.

  Also promising for the measurement of $\alpha$
is the decay $B^0 \to a_0^\pm(980) \pi^\mp$, which is expected
to be free of main tree contributions.
The analysis relies on a neural network algorithm to select
$B^0$ decays to $a_0^\pm(980) \pi^\mp$, with $a_0^\pm \to \eta \pi^\pm$
and $\eta \to \gamma\gamma$.
A likelihood fit using the neural network output, $\gamma\gamma$
mass, $m_{ES}$ and $\Delta E$ extracts a signal yield of
$18.1^{+8.7}_{-7.4}$ events with a corresponding statistical
significance of $3.7\:\sigma$.
The $\eta \pi^\pm$ invariant mass is not included in the fit
because the parameters of the $a_0$ resonance are not well known.

  Finally, we also measured the branching fraction for $B^+$
decays into $K^{\ast 0} \pi^+$. Part of the interest in this
process stems from the expectation that tree and penguin amplitudes
have comparable magnitudes, which makes this decay an ideal candidate
for the observation of direct CP violation, see Eq.~\ref{equ:acptp}.
The measured yields and branching fractions for this and all other
rare hadronic decays are summarized in Table~\ref{tab:bfrachad}.

\section{DIRECT CP VIOLATION}

  We searched for CP violation in the decay of the
flavor self-tagged modes listed in Table~\ref{tab:cpv}.
The table shows the measured CP-asymmetries as defined
in Eq.~\ref{equ:acpdef}.
For the $B^0 \to K^+ \pi^-$ decay mode, an on-resonance
sample with an integrated luminosity of 30.4~\invfb\
is used~\cite{2bcpv}.
In the case of the $B^0 \to \rho^\pm \pi^\mp$ mode,
no tagging of the $B^0$ flavor is attempted and the
asymmetry quoted in Table~\ref{tab:cpv}
corresponds to the asymmetry between the number of
$\rho^+\pi^-$ and $\rho^-\pi^+$ events.
No evidence for direct CP violation has been found but
the precision is now better than 10\% in some of the
more abundant modes and, thus, these data become more useful
in constraining models.

\begin{table}[tbp]
\caption{Summary of direct CP violation measurements.
         The first and second uncertainties
         are statistical and systematic, respectively.
}
\label{tab:cpv}
\begin{center}
\begin{tabular}{ll}
Mode              & ~~~~~~~~~~~~$A_{CP}$ \\
\hline			   
$\Kstar \gamma$~\cite{KstrGam}   & $-0.044 \pm 0.076 \pm 0.012$ \\ 
$K^+\pi^-$~\cite{2bcpv}          & $-0.07  \pm 0.08  \pm 0.04$ \\ 
$K^+\pi^0$~\cite{2body}          & ~~$0.00  \pm 0.18  \pm 0.04$ \\ 
$K^0\pi^+$~\cite{2body}          & $-0.21  \pm 0.18  \pm 0.03$ \\ 
$\phi K^+$~\cite{phiK}           & $-0.05  \pm 0.20  \pm 0.03$ \\ 
$\phi \Kstarp$~\cite{phiK}       & $-0.43^{+0.36}_{-0.30} \pm 0.06$ \\ 
$\phi \Kstarz$~\cite{phiK}       & ~~$0.00  \pm 0.27  \pm 0.03$ \\ 
$\omega\pi^+$~\cite{q2bcpv}      & $-0.01^{+0.29}_{-0.31} \pm 0.03$ \\ 
$\etap K^+$~\cite{q2bcpv}        & $-0.11  \pm 0.11  \pm 0.02$ \\ 
$\rho^\pm\pi^\mp$~\cite{3pi}     & $-0.04  \pm 0.18  \pm 0.02$ \\ 
\hline
\end{tabular}
\end{center}
\end{table}

\section{SUMMARY AND OUTLOOK}

  The large data sample accumulated at the SLAC PEP-II B Factory
is already providing a substantial amount of interesting
results. We are now sensitive to rare \B\ decays with branching
fractions as low as $4 \times 10^{-6}$. A search for direct CP violation
in radiative penguin and rare hadronic decays finds no evidence
for CP violation in the decay process.
It should be noted that some of the results presented here are preliminary.

  PEP-II is expected to deliver an integrated luminosity
of approximately 100 \invfb\ by the summer of 2002,
which will provide a five-fold increase in statistics compared
to the results reported here.
This will allow many more rare processes to be studied.


\begin{thebibliography}{99}

\bibitem{sin2b} B. Aubert {\it et al.}, Phys. Rev. Lett. 87, 091801 (2001);
                K. Abe {\it et al.}, Phys. Rev. Lett. 87, 091802 (2001).
\bibitem{bbrdet}  B. Aubert {\it et al.}, SLAC-PUB-8569, to appear
                  in Nucl. Instr. and Methods.
\bibitem{KstrGam} B. Aubert {\it et al.}, SLAC-PUB-8952, hep-ex/0110065.
\bibitem{GamGam}  B. Aubert {\it et al.}, Phys. Rev. Lett. 87, 241803 (2001).
\bibitem{Kll}     B. Aubert {\it et al.}, SLAC-PUB-8910, hep-ex/0107026.
\bibitem{2body}   B. Aubert {\it et al.}, Phys. Rev. Lett. 87, 151802 (2001).
\bibitem{kzkzb}   B. Aubert {\it et al.}, SLAC-PUB-8978, hep-ex/0109005.
\bibitem{phiK}    B. Aubert {\it et al.}, Phys. Rev. Lett. 87, 151801 (2001).
\bibitem{etapom}  B. Aubert {\it et al.}, Phys. Rev. Lett. 87, 221802 (2001).
\bibitem{etak}    B. Aubert {\it et al.}, SLAC-PUB-8914, hep-ex/0107037.
\bibitem{etapX}   B. Aubert {\it et al.}, SLAC-PUB-8979, hep-ex/0109034.
\bibitem{a0pi}    B. Aubert {\it et al.}, SLAC-PUB-8930, hep-ex/0107075.
\bibitem{3pi}     B. Aubert {\it et al.}, SLAC-PUB-8926, hep-ex/0107058.
\bibitem{Kstrpi}  B. Aubert {\it et al.}, SLAC-PUB-8981, hep-ex/0109007.
\bibitem{2bcpv}   B. Aubert {\it et al.}, SLAC-PUB-8929, hep-ex/0107074.
\bibitem{q2bcpv}  B. Aubert {\it et al.}, SLAC-PUB-8980, hep-ex/0109006.

\end{thebibliography}
\end{document}